\documentclass[prl,aps,amsmath,showpacs,superscriptaddress,twocolumn]{revtex4-1}
\usepackage{graphicx}
\usepackage{dcolumn}
\usepackage{bm}
\usepackage{subfigure}
\usepackage{xspace}
\usepackage{textpos}
\usepackage[english]{babel}

\usepackage{amsmath}
\usepackage[mathlines]{lineno}
\usepackage{color}
\usepackage[bookmarksnumbered, pdfstartview=FitH,colorlinks,citecolor=blue,linkcolor=blue,]{hyperref}

\usepackage[perpage,symbol]{footmisc}
\usepackage{indentfirst}
\usepackage{fancyvrb}
\usepackage{multirow}
\usepackage{float}
\newcommand{\ra}{\rightarrow}

\newcommand{\jpsi}{J/\psi}
\newcommand{\pio}{\pi^{0}}
\newcommand{\pip}{\pi^{+}}
\newcommand{\pim}{\pi^{-}}
\newcommand{\etap}{\eta^{\prime}}
\newcommand{\chisq}{\chi^{2}}

\newcommand{\rhop}{\rho^{+}}
\newcommand{\rhom}{\rho^{-}}
\renewcommand{\thefootnote}{\fnsymbol{footnote}}

\begin{document}


\title{\bf Amplitude Analysis of the Decays $\etap \ra \pip\pim\pio$ and $\etap\ra \pio\pio\pio$}
  
\author{
  \begin{small}
    \begin{center}
  M.~Ablikim$^{1}$, M.~N.~Achasov$^{9,e}$, X.~C.~Ai$^{1}$,
  O.~Albayrak$^{5}$, M.~Albrecht$^{4}$, D.~J.~Ambrose$^{44}$,
  A.~Amoroso$^{49A,49C}$, F.~F.~An$^{1}$, Q.~An$^{46,a}$,
  J.~Z.~Bai$^{1}$
% ...
  \\
  \vspace{0.2cm}
  (BESIII Collaboration)\\
  \vspace{0.2cm} {\it
    $^{1}$ Institute of High Energy Physics, Beijing 100049, People's Republic of China\\
% ...
$^{4}$ Bochum Ruhr-University, D-44780 Bochum, Germany\\
    $^{5}$ Carnegie Mellon University, Pittsburgh, Pennsylvania 15213, USA\\
% ...
$^{9}$ G.I. Budker Institute of Nuclear Physics SB RAS (BINP), Novosibirsk 630090, Russia\\
% ...
$^{44}$ University of Rochester, Rochester, New York 14627, USA\\
% ...
$^{46}$ University of Science and Technology of China, Hefei 230026, People's Republic of China\\
% ...
$^{49}$ (A)University of Turin, I-10125, Turin, Italy; (B)University of Eastern Piedmont, I-15121, Alessandria, Italy; (C)INFN, I-10125, Turin, Italy\\
% ...
$^{a}$ Also at State Key Laboratory of Particle Detection and Electronics, Beijing 100049, Hefei 230026, People's Republic of China\\
% ...
$^{e}$ Also at the Novosibirsk State University, Novosibirsk, 630090, Russia\\
% ...
  }
    \end{center}
    \vspace{0.4cm}
  \end{small}
}

\begin{abstract}
Based on a sample of $1.31\times10^9$ $\jpsi$ events collected with
the BESIII detector, an amplitude analysis of the isospin-violating
decays $\etap\ra\pi^+\pi^-\pi^0$ and $\etap\ra\pi^0\pi^0\pi^0$ is
performed.  A significant $P$-wave contribution from
$\etap\ra\rho^{\pm}\pi^{\mp}$ is observed for the first time in
$\etap\ra\pip\pim\pio$.  The branching fraction is determined to be
${\mathcal
  B}(\etap\ra\rho^{\pm}\pi^{\mp})=(7.44\pm0.60\pm1.26\pm1.84)\times
10^{-4}$, where the first uncertainty is statistical, the second
systematic, and the third model dependent.  In addition to the
nonresonant $S$-wave component, there is a significant $\sigma$
meson component.  The branching fractions of the combined $S$-wave
components are determined to be ${\mathcal
  B}(\etap\ra\pip\pim\pio)_S=(37.63\pm0.77\pm2.22\pm4.48)\times
10^{-4}$ and ${\mathcal
  B}(\etap\ra\pio\pio\pio)=(35.22\pm0.82\pm2.54)\times 10^{-4}$,
respectively.  The latter one is consistent with previous BESIII
measurements.

\end{abstract}


\maketitle

The decays $\etap\ra\pi\pi\pi$ are isospin-violating processes. Because
the electromagnetic contribution is strongly
suppressed~\cite{Sutherland1966384, Baur1996127}, they are induced
dominantly by the strong interaction via the explicit breaking of
chiral symmetry by the $d-u$ quark mass difference.  In recent years,
there has been considerable interest in these decays because they
allow the determination of the light quark mass difference using the
ratios of decay widths,
$r_{\pm}=\mathcal{B}(\etap\ra\pi^+\pi^-\pi^0)/\mathcal{B}(\etap\ra\pi^+\pi^-\eta)$
and
$r_{0}=\mathcal{B}(\etap\ra\pi^0\pi^0\pi^0)/\mathcal{B}(\etap\ra\pi^0\pi^0\eta)$~\cite{PhysRevD192188,
  Borasoy200641}.  Within the framework of chiral effective field
theory combined with a relativistic coupled-channel approach,
Ref.~\cite{Borasoy2005383} predicts that the
$\eta^\prime\rightarrow\rho^\pm\pi^\mp$ $P$-wave contribution should
be large for $\etap\ra\pip\pim\pio$.  For the channel with three
neutral pions, $\etap\ra\pio\pio\pio$, the $P$-wave contribution in
two-body rescattering is forbidden by Bose symmetry. In general, the
final-state interaction is expected to be very important because it was
already found to be essential to explain the decay width of
$\eta\ra\pi\pi\pi$~\cite{Gasser1985539,Bijnens2007030}. In the case
of $\etap$ decays, the final-state interaction is further enhanced due
to the presence of nearby resonances and is expected to strongly
affect the values of the branching fractions and the Dalitz plot
distributions.

So far, there is no direct experimental evidence of an intermediate
$\rho^\pm$ contribution to the decay $\etap\ra\pip\pim\pio$.  In 2009,
the CLEO-c experiment~\cite{Naik2009061801} reported the first
observation of $\etap\ra\pip\pim\pio$ with $20.2^{+6.1}_{-4.8}$
events, corresponding to a branching fraction of
$(37\pm11)\times10^{-4}$, and a Dalitz plot consistent with a flat
distribution. Recently the decay was also observed by the BESIII
experiment~\cite{Ablikim2012182001} with a branching fraction
consistent with the CLEO-c result; however, no Dalitz plot analysis
was presented.  Interest in the decay channel $\eta^\prime\ra
\pi^0\pi^0\pi^0$ stems from the observed 4$\sigma$ discrepancy between
the recent branching fraction measurement by BESIII
[$(35.6\pm4.0)\times10^{-4}$]~\cite{Ablikim2012182001} and those from
all previous experiments~\cite{Binon1984264,Alde1987603,Blik20082124}.
The BESIII result indicates a value for the ratio
$r_0$ that is two times larger than previous experiments. Furthermore, the recent determination of the Dalitz plot
slope parameter for $\eta^\prime\rightarrow\pi^0\pi^0\pi^0$ decay
gave $\alpha=-0.687\pm0.061$~\cite{PhysRevD92012014}, which deviates
significantly from that for the phase-space distribution ($\alpha=0$).
This implies that final-state interactions play an essential role.
In this Letter, we present
an amplitude analysis combining
$\etap\ra\pip\pim\pio$ and $\etap\ra \pio\pio\pio$ events originating
from $J/\psi$ radiative decays using $1.31 \times 10^9$ $J/\psi$
events~\cite{Yanghx2014, Yanghxnew} accumulated by the BESIII detector,
which is described in detail in Ref.~\cite{Ablikim2010345}.

For a $\jpsi\ra\gamma\etap$ with
$\etap\ra\pip\pim\pio$ candidate event, two tracks with opposite
charge and at least three photon candidates are required.
The selection criteria for charged tracks and photon candidates are the same as those in Ref.~\cite{PhysRevD92012014}.
Because the radiative photon from the $\jpsi$ is always more energetic than the photons from the $\pio$ decays,
the photon candidate with the maximum energy in the event
is taken as the radiative one.
For each $\pip\pim\gamma\gamma\gamma$ combination,
a six-constraint (6$C$) kinematic fit is performed, and the $\chi^{2}_{6C}$ is required to be less than 25.
The fit enforces energy-momentum conservation and constrains the invariant masses of 
the other photon pair and $\pip\pim\pio$
to the nominal $\pio$ and $\etap$ mass, respectively.
If there are more than three photon candidates
in an event, the combination with the smallest $\chi^{2}_{6C}$ is retained.
To reject possible backgrounds with two or four photons in the final states,
we further require that the probability of the 4$C$ kinematic fit imposing energy-momentum conservation for
the $J/\psi\ra\pip\pim\gamma\gamma\gamma$ signal hypothesis is larger than that for the
$J/\psi\ra\pip\pim\gamma\gamma$ and $J/\psi\ra\pip\pim\gamma\gamma\gamma\gamma\gamma$ background hypotheses.
Additionally, events with $|M(\gamma\pio)-m_{\omega}| < 0.05\;\text{GeV}/c^{2}$ are rejected
to suppress background from $\jpsi\ra\omega\pip\pim$. 

\begin{figure}[!htbp]
    \centering
    \includegraphics[width=0.23\textwidth]{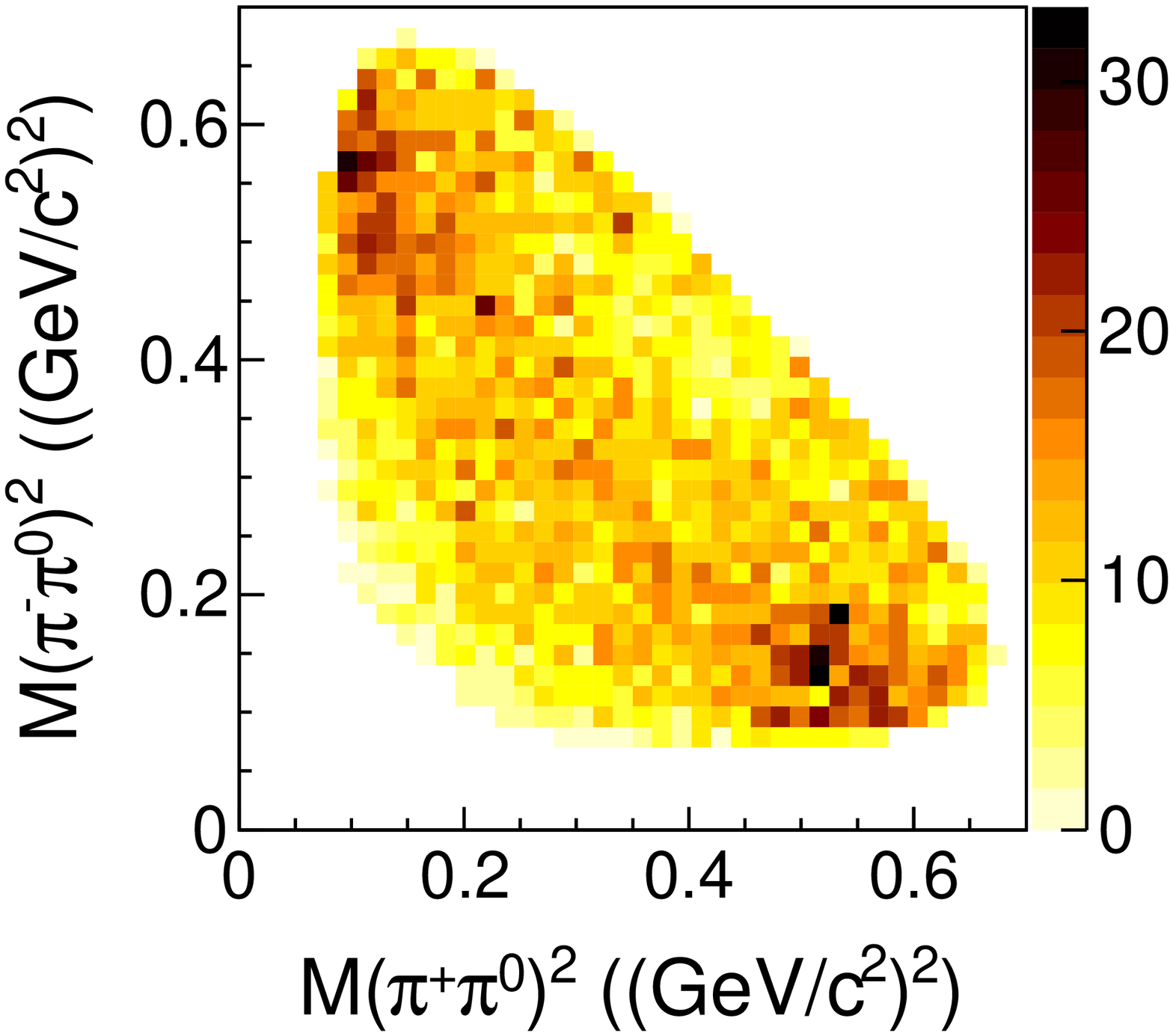}\put(-40,90){\bf (a)}
    \includegraphics[width=0.23\textwidth]{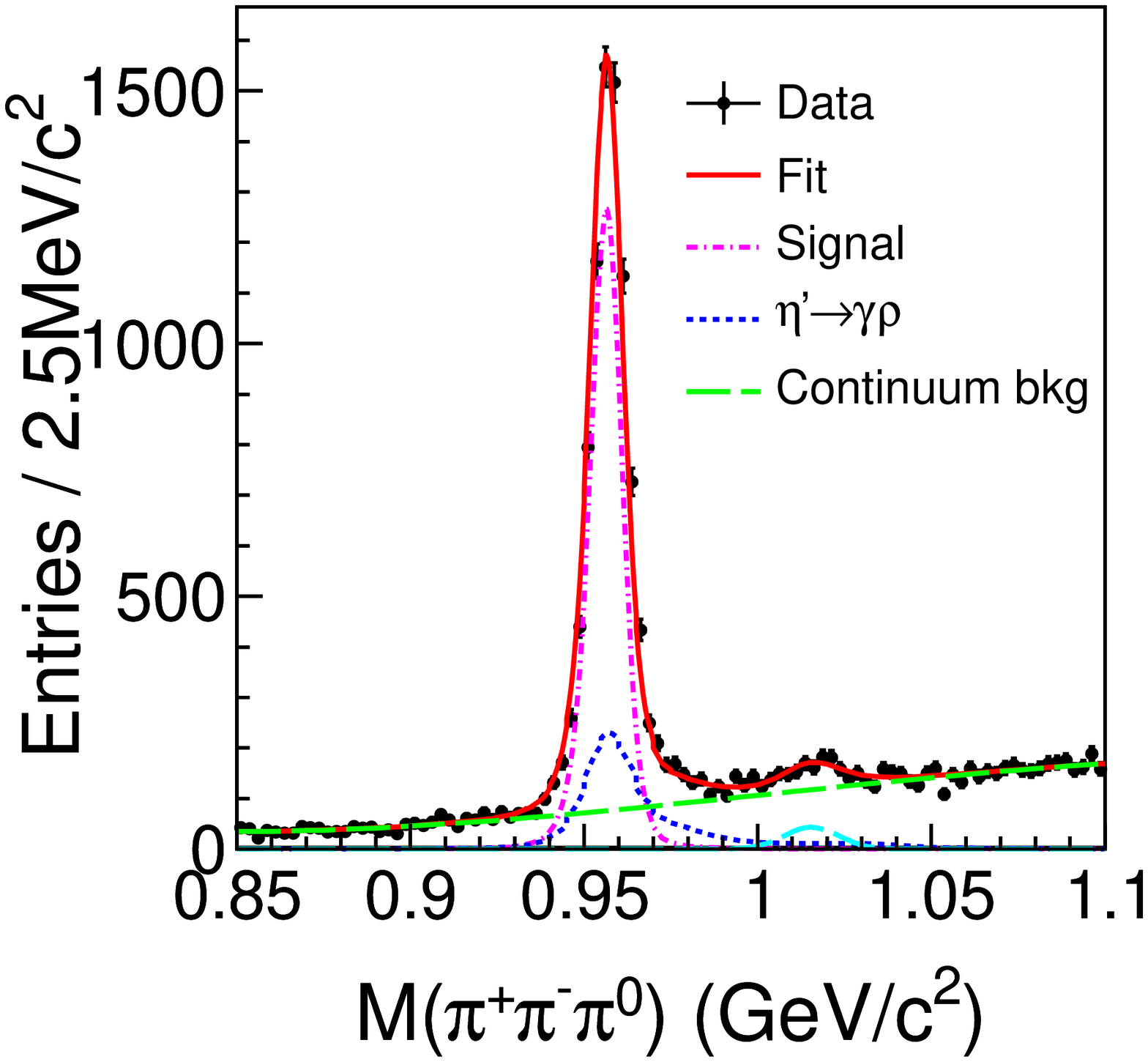}\put(-85,90){\bf (b)}
    \caption{\label{fig:chadp_data} (a) $\etap\ra\pip\pim\pio$ Dalitz
      plot for candidate events selected from data.  (b) Invariant mass
      distribution of $\pip\pim\pio$ candidates without the $\etap$ mass
      constraint applied in the kinematic fit.}
\end{figure}

With the above requirements, a sample of 8267 events is selected, and
the corresponding Dalitz plot is shown in
Fig.~\ref{fig:chadp_data}~(a), where two clusters of events
corresponding to the decays of $\etap\ra\rho^\pm\pi^\mp$ are observed.
The possible background events are investigated with a (Monte Carlo) MC sample of
$1.2\times 10^9 J/\psi$ inclusive decays generated with the
LUNDCHARM and EVTGEN models~\cite{lundcharm, evtgen}.  Using
the same selection criteria, the surviving background events mainly
originate from the decay $\etap\ra\gamma\rho$ with $\rho\ra\pi\pi$ or
$\rho\ra\gamma\pi\pi$, which accumulate in a peak around the
$\eta^\prime$ mass region, and the nonpeaking processes with
multiphotons in the final states, e.g.,
$J/\psi\ra\pi^+\pi^-\pi^0\pi^0$. However, none of these backgrounds
contribute to the clusters around the $\rho^\pm$ mass region.  For
$\etap\ra\gamma\rho$, a study with a dedicated MC simulation based on
an amplitude analysis of the same BESIII data and Ref.~\cite{ToledoPRD76033001} and using the
branching fractions of $J/\psi\ra\gamma\eta^\prime$ and
$\etap\ra\gamma\rho$,
  $\rho\ra\pi\pi/\gamma\pi\pi$,
  $\pio\ra\gamma\gamma$~\cite{PDG2014090001} predicts the
number of events from this background to be $1362\pm54$.

The decay $J/\psi\ra\pi^+\pi^-\pi^0\pi^0$, which is assumed to
represent the nonpeaking background contribution, is not well known.
In order to estimate this background, an
alternative data sample is selected by using a 5$C$ kinematic fit
without the $\etap$ mass constraint.  The resulting $\pi^+\pi^-\pi^0$
invariant mass spectrum is shown in Fig.~\ref{fig:chadp_data} (b),
where the $\etap$ peak is clearly visible. We then perform an unbinned
maximum likelihood fit to the $M(\pip\pim\pio)$ distribution where the
signal is described by the MC simulated shape convolved with a
Gaussian resolution function, the peaking background
($\eta^\prime\ra\gamma\rho$) is described by the MC simulated shape,
and the nonpeaking background contribution by a second-order
Chebyshev polynomial function.  The number of $\etap\ra\gamma\rho$
events is fixed to the expected value, while the small peak around 1.02
GeV/$c^2$ from $J/\psi\ra\gamma\gamma\phi$ events is described with a
Gaussian function.  The number of nonpeaking background events in the
selected 6$C$-fitted sample is estimated to be $838\pm31$, using the
number of background events from the 5$C$-fitted sample in the $\etap$
signal region $(|M(\pip\pim\pio)-0.958|<0.02$GeV/c$^{2})$
and taking into account the slight difference of
detection efficiency between 5$C$ and 6$C$ kinematic requirements.
To further verify the above background estimation,
we checked the background shapes in $\pi\pi$ mass spectra of the data.
For each mass bin, the number of background events is extracted by fitting the
$\pip\pim\pio$ mass spectrum in this bin. We found that the background
shapes are consistent with those estimated from the MC simulations.
(More details are given in the Supplemental Material~\cite{supplemental}.)

For $\jpsi\ra\gamma\etap$ with $\etap\ra\pio\pio\pio$, events
containing at least seven photon candidates and no charged tracks are
selected.  The photon selection criteria are the same as those for
$\etap\ra\pip\pim\pio$.  The photon with the maximum energy in the
event is assumed to be the radiative photon originating from the decay
of $\jpsi$.  For the remaining photon candidates, pairs of photons are
combined to form $\pio\ra\gamma\gamma$ candidates which are subjected
to a 1$C$ kinematic fit, where the invariant mass of the photon pair is
constrained to the nominal $\pio$ mass, and the $\chisq$ value is
required to be less than 25.  To suppress $\pio$ miscombinations, the
$\pio$ decay angle ($\theta_{\rm decay}$), defined as the polar angle
of a photon in the corresponding $\gamma\gamma$ rest frame, is
required to satisfy $|\cos \theta_{\rm decay}|<0.95$.  From the
accepted $\pio$ candidates and the corresponding radiative photon,
$\gamma\pio\pio\pio$ combinations are formed.  A kinematic fit with
eight constraints (8$C$) is performed, constraining the invariant masses
of $\gamma\gamma$ pairs and $\pio\pio\pio$ candidates to the nominal
$\pio$ and $\etap$ masses, respectively.  Events with $\chi^2_{8C}<70$
are accepted for further analysis.  If there is more than one
combination, only the one with the smallest $\chisq_{8C}$ is retained.
To suppress possible background from $\jpsi\ra\gamma\eta\pio\pio$, a
7$C$ kinematic fit is performed under the $\jpsi\ra\gamma\eta\pio\pio$
hypothesis and events for which the probability of this 7$C$ fit is
larger than that of the signal hypothesis are discarded.  In addition,
events which have at least one $\gamma\gamma$ pair with invariant mass
within the $\eta$ signal region, $(0.52,0.59)$ GeV/c$^{2}$, are
rejected.  Possible background from $\jpsi\ra\omega\pio\pio$ is
suppressed by vetoing events with $|M(\gamma\pio)-m_{\omega}|<0.05$
GeV/c$^{2}$, where $M(\gamma\pio)$ is the invariant mass of a
$\gamma\pio$ combination.

\begin{figure}[!htbp]
    \centering
		\includegraphics[width=0.23\textwidth]{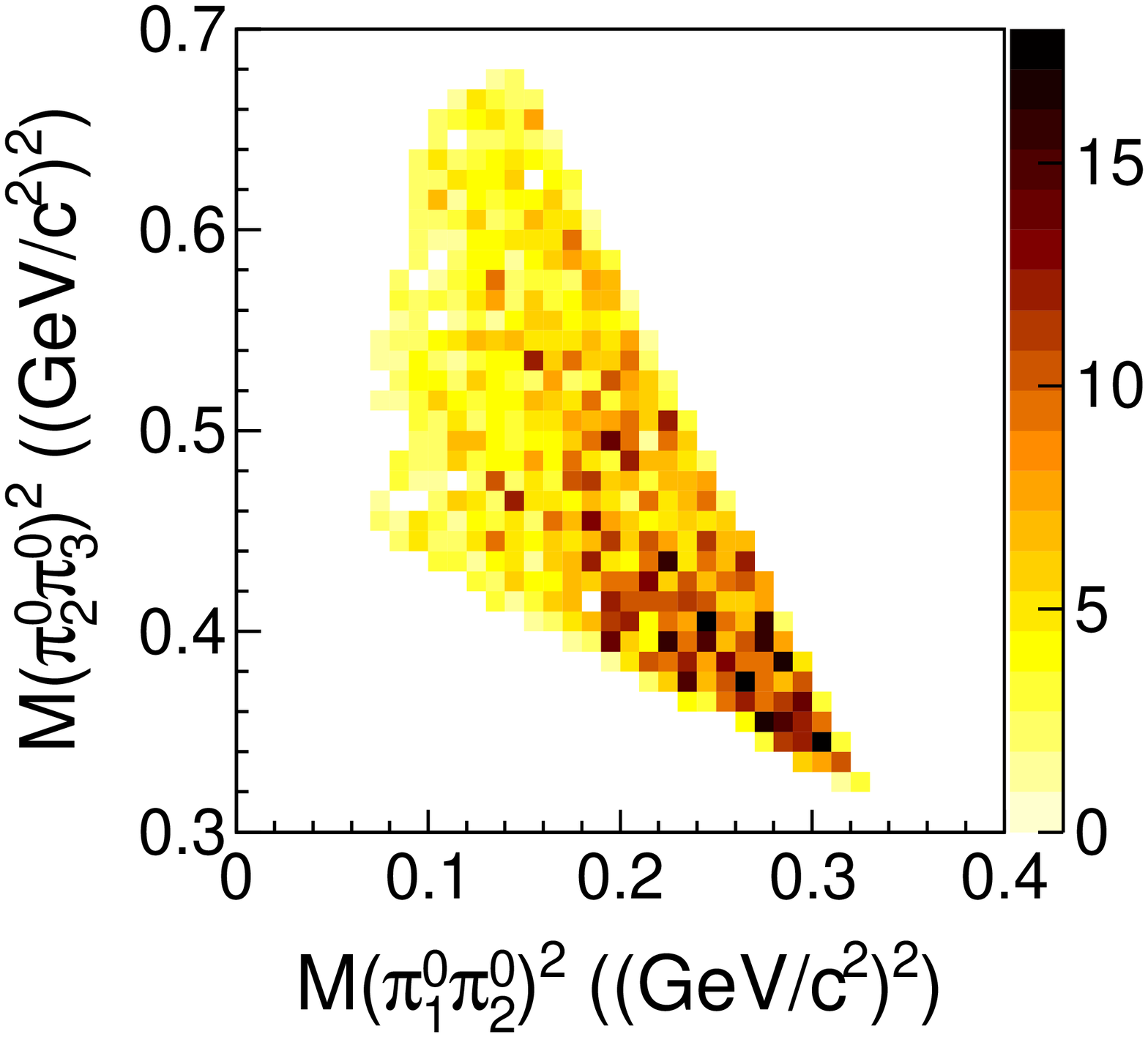}\put(-40,90){\bf (a)}
		\includegraphics[width=0.23\textwidth]{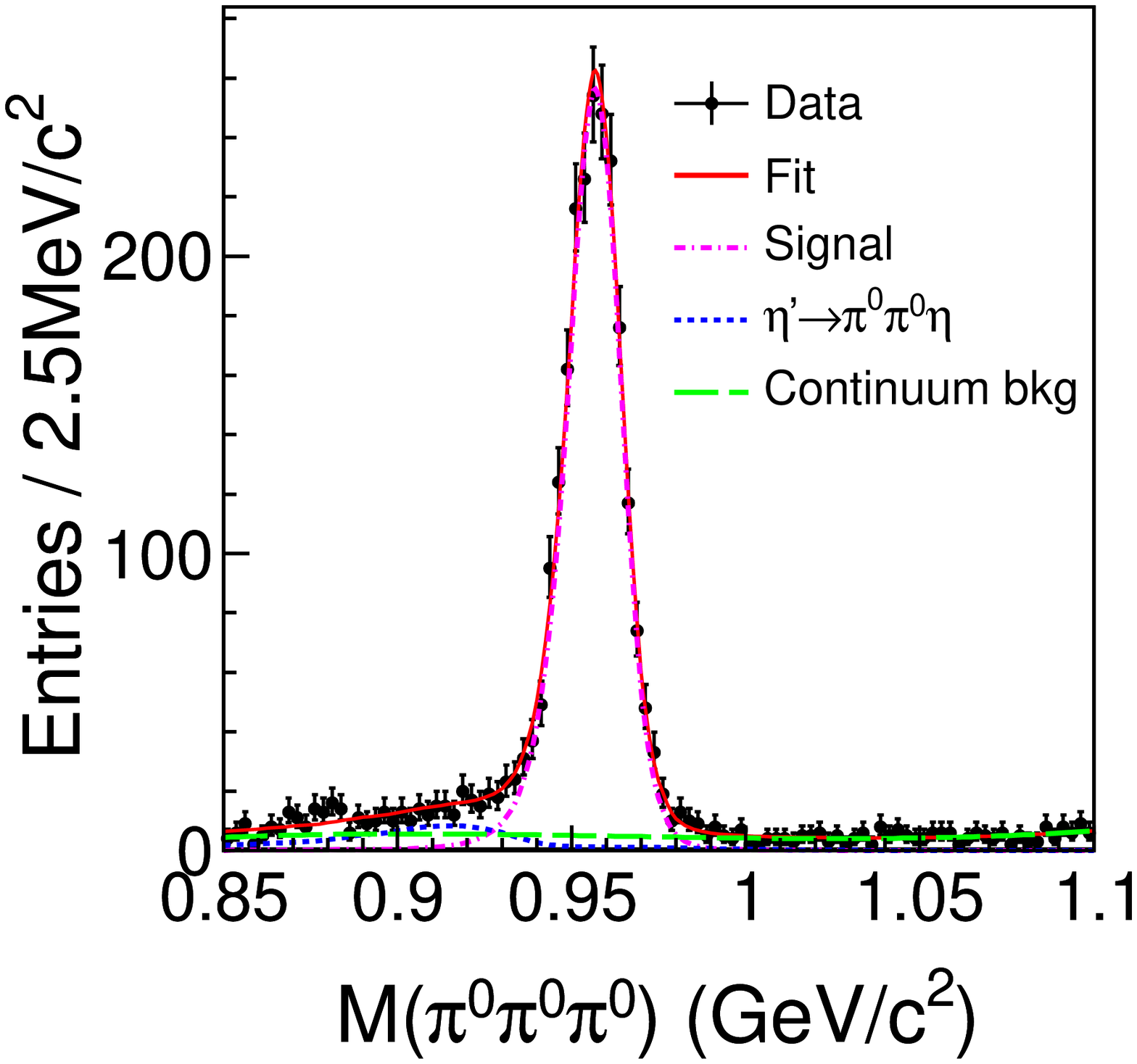}\put(-85,90){\bf (b)}
    \caption{\label{fig:neudp_data} (a)
    $\etap\ra\pio\pio\pio$ Dalitz plot for candidate
    events selected from data.  (b) Invariant mass of
    $\pio\pio\pio$ candidates without the $\etap$ mass
    constraint applied in the kinematic fit.}
\end{figure}

The three $\pi^0$ candidates selected are ordered as $\pi^0_1$,
$\pi^0_2$, and $\pi^0_3$ according to their descending energies in the
$\etap$ rest frame, and the corresponding Dalitz plot is displayed in
Fig.~\ref{fig:neudp_data} (a) for the 2237 events selected.  The
analysis of the inclusive MC sample of $1.2\times 10^9$ $\jpsi$ decays
indicates a low background level, including the peaking background
originating from $\jpsi\ra\gamma\etap$ with
$\etap\ra\eta\pio\pio$ and the nonpeaking background mainly coming
from $J/\psi\ra\gamma\pio\pio\pio$, since the decay of
$J/\psi\ra\pi^0\pi^0\pi^0\pi^0$ is forbidden.  The number of
background events from $\etap\ra\eta\pio\pio$ is estimated to be
$46\pm3$, using a MC sample with the decay amplitudes from
Ref.~\cite{Blik2009231}.  Similarly, we perform a 7$C$ kinematic fit
without applying the constraint on the $\eta^\prime$ mass to estimate
the nonpeaking background.  The fit to the $M(\pio\pio\pio)$ distribution
is displayed in Fig.~\ref{fig:neudp_data} (b) using the simulated
shape convolved with a Gaussian resolution function for the signal, a
MC simulated peaking background shape, and a second-order polynomial
function for nonpeaking background events.  The number
of the nonpeaking background events in the selected $\etap\ra
\pio\pio\pio$ sample, predominantly originating from
$J/\psi\ra\gamma\pio\pio\pio$, is estimated to be $176\pm24$ after
taking into account the detection efficiencies with and without the
$\etap$ mass constraint.

A Dalitz plot analysis based on the formalism of the isobar
model~\cite{PRD89052001} is performed.  The resonant $\pi$-$\pi$
$S$-wave ($L=0$ for $\sigma$) and $P$-wave ($L=1$ for $\rho^{\pm}$)
amplitudes are described following the formalism from
Ref.~\cite{PRD83074004},
\begin{equation}
W(s)  =  \frac{1}{\cot\delta_{L}(s)-i},
\end{equation}
where
\[\cot\delta_{0}(s) = \frac{\sqrt{s}}{2k}\frac{M_{\pi}^{2}}{s-M_{\pi}^{2}/2}
\left\{\frac{M_{\pi}}{\sqrt{s}} + B_{0}^{S} + B_{1}^{S}\omega_{0}(s)\right\},
\]
\[
\cot\delta_{1}(s) = \frac{\sqrt{s}}{2k^{3}}\left(M_{\rho}^{2}-s\right)
\left\{\frac{2M_{\pi}^{3}}{M_{\rho}^{2}\sqrt{s}} + B_{0}^{P} + B_{1}^{P}
\omega_{1}(s)\right\},
\]
\[
\omega_{L}(s) = \frac{\sqrt{s}-\sqrt{s_{L}-s}}{\sqrt{s}+\sqrt{s_{L}-s}}-1.
\]
Here $s$ is the $\pi\pi$ invariant mass square, 
$k=\sqrt{s/4-M^2_{\pi}}$, $\sqrt{s_{0}}=2M_{K}$, the masses
$M_{\rho}$, $M_{K}$, and $M_{\pi}$ are fixed to the world average
values~\cite{PDG2014090001}, $\sqrt{s_{1}}=1.05$ GeV
is a constant, and $B_{0}^{S}$, $B_{1}^{S}$, $B_{0}^{P}$, and
$B_{1}^{P}$ are free parameters.

The free parameters of the probability density function (PDF) are
optimized with an unbinned maximum likelihood fit using both the
$\etap\ra\pip\pim\pio$ and $\etap\ra\pio\pio\pio$ events, where the
background contributions are included as noninterfering terms in the
PDF and are fixed according to the MC simulation, the mass resolution, and the detection efficiency
obtained from the MC simulation are taken
into account in the signal PDF.  The fit minimizes the negative
log-likelihood value $-\ln{\mathcal {L}} = -\sum_{i=1}^{N_{1}}\ln
\mathcal{P}_{i}-\sum_{j=1}^{N_{2}}\ln \mathcal{P}^{\prime}_{j}$, where
$\mathcal{P}_{i}$ and $\mathcal{P}^{\prime}_{j}$ are the PDFs for an
$\etap\ra\pip\pim\pio$ event $i$ and an $\etap\ra\pio\pio\pio$ event
$j$, respectively.  The sum runs over all accepted events.  From
charge conjugation invariance, the magnitude and phase for $\rhop$ and
$\rhom$ are taken to be the same in the nominal fit.

\begin{figure}[!htbp]
    \centering
    \includegraphics[width=0.23\textwidth]{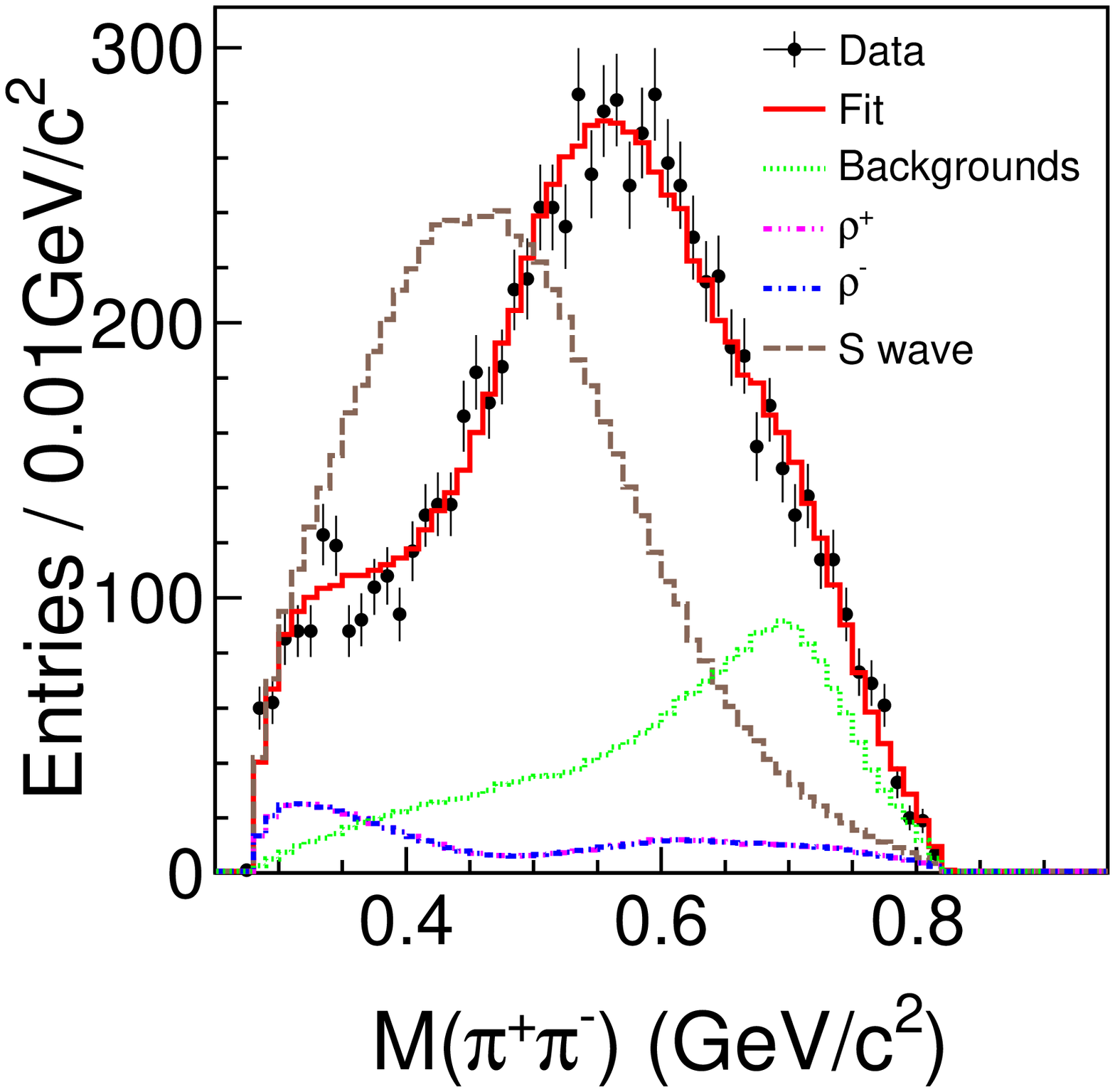}\put(-85,90){\bf (a)}
    \includegraphics[width=0.23\textwidth]{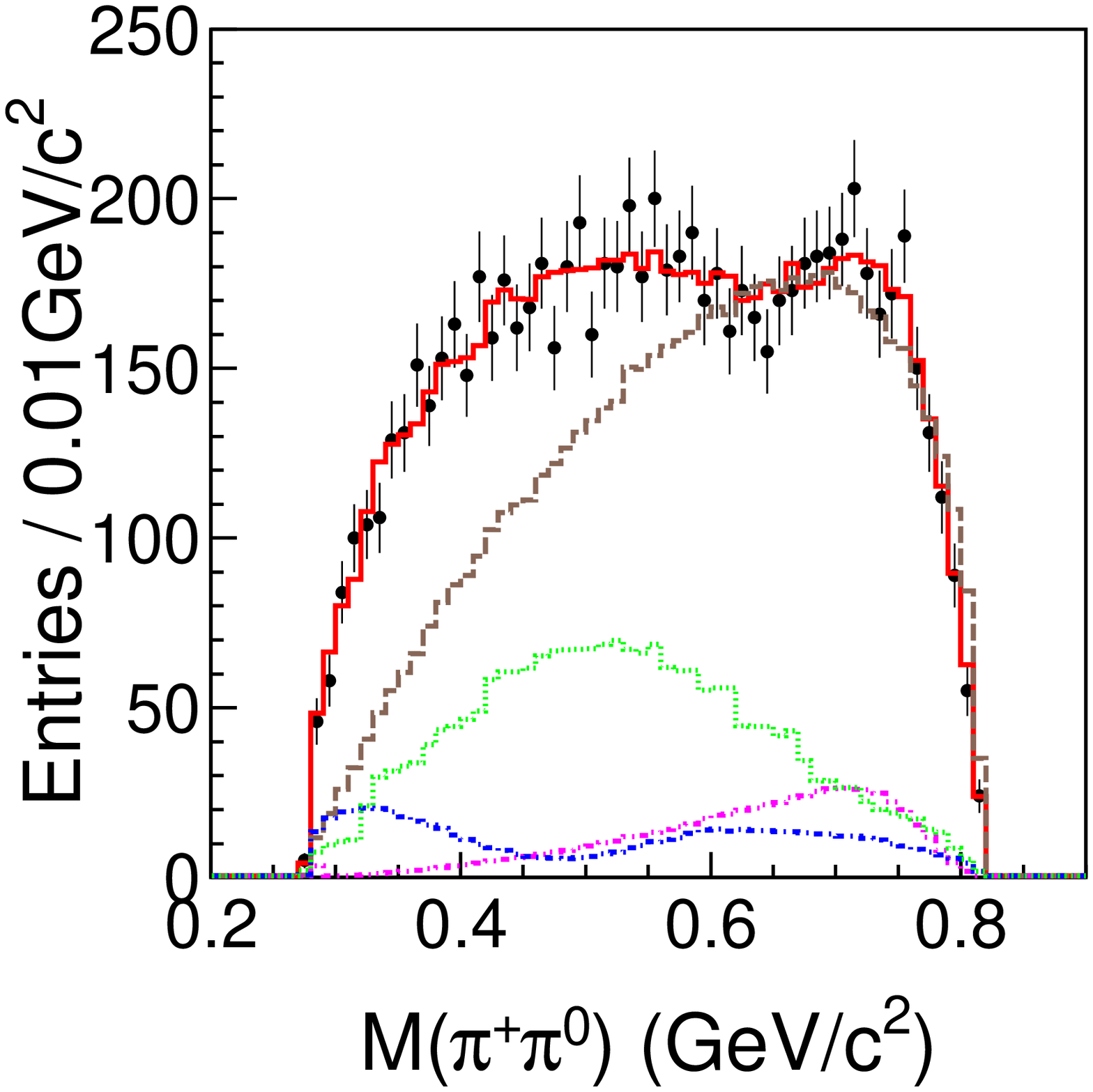}\put(-85,90){\bf (b)}

    \includegraphics[width=0.23\textwidth]{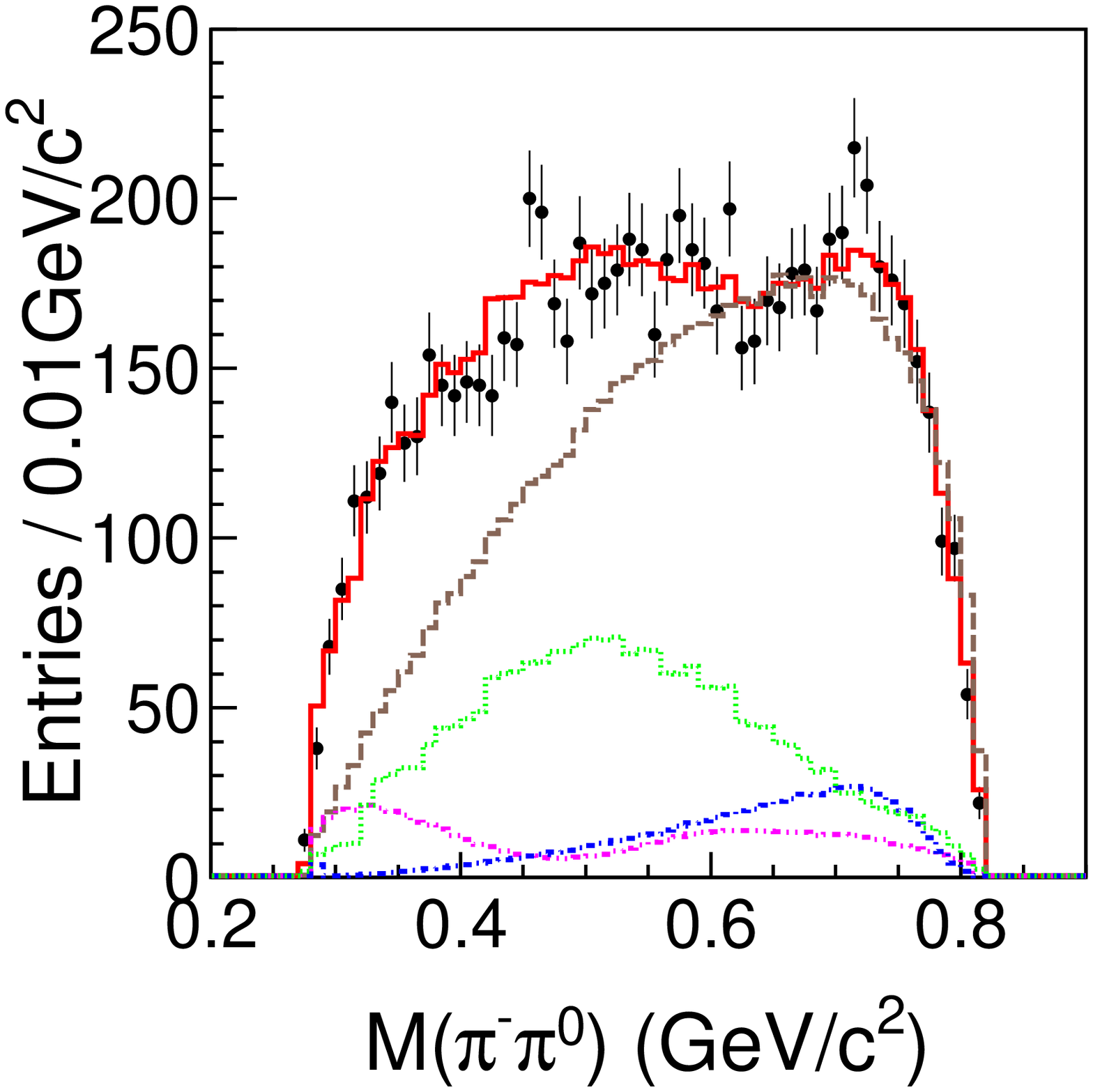}\put(-85,90){\bf (c)}
    \includegraphics[width=0.23\textwidth]{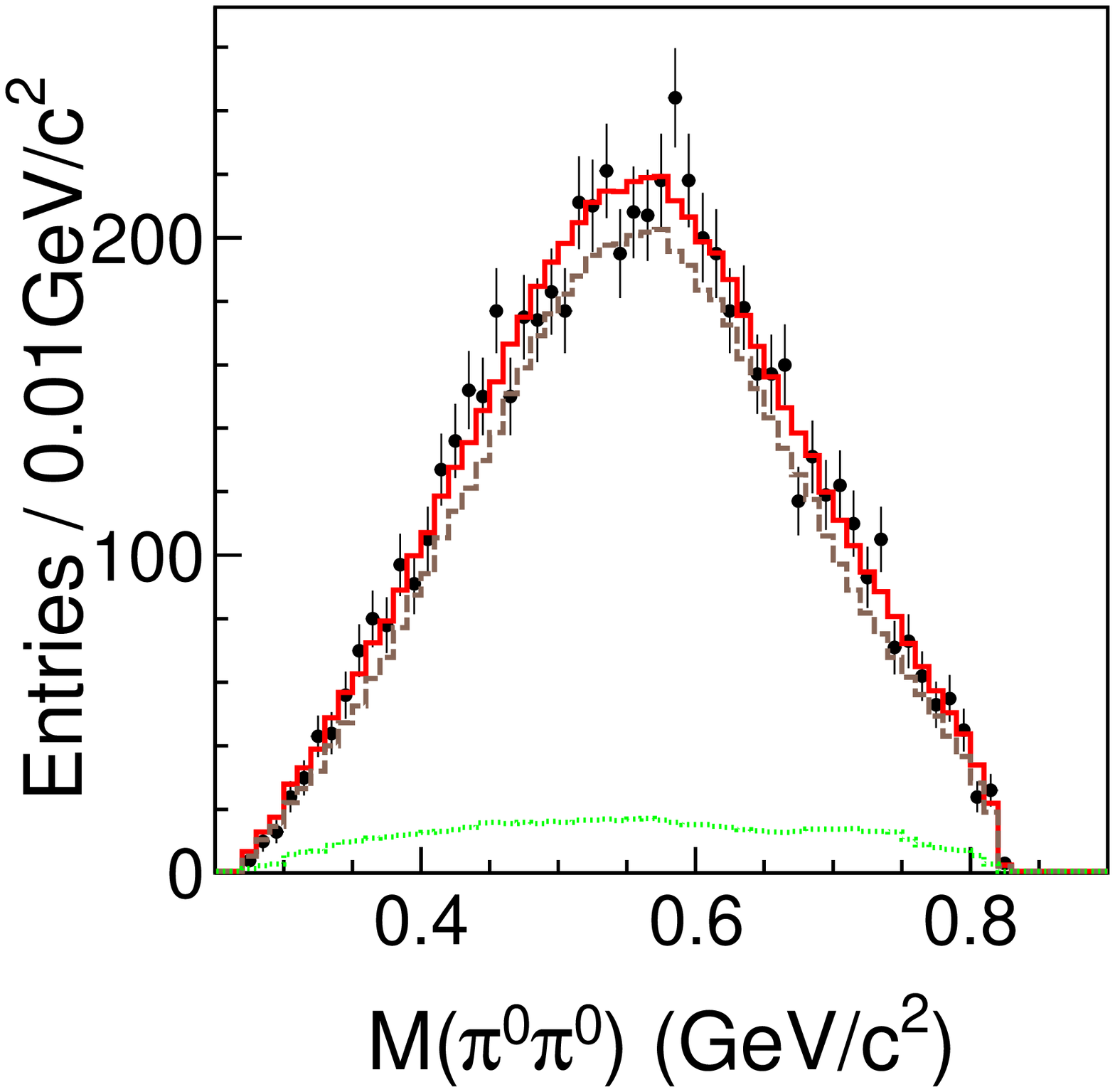}      \put(-85,90){\bf (d)}
   \caption{\label{fig:fitres} Comparison of the invariant mass distributions of (a) $\pip\pim$,
	 (b) $\pip\pio$, (c) $\pim\pio$, and (d) $\pio\pio$ between data (dots with error bars)
	 and the fit result projections (solid histograms).
	 The dotted, dashed, dash-dotted, and dash-dot-dotted histograms show
	 the contributions from background, $S$ wave, $\rhom$, and $\rhop$, respectively.
		}
\end{figure}

Projections of the data and fit results are displayed in
Fig.~\ref{fig:fitres}. The data are well described by three
components: $P$ wave ($\rho^\pm\pi^\mp$), resonant $S$ wave
($\sigma\pi^0$), and phase-space $S$ wave ($\pi\pi\pi$).  The
interference between $\sigma$ and the nonresonant term is large and
strongly depends on the parametrization of $\sigma$. Therefore we are
unable to determine the individual contributions and consider only the
sum of the $S$-wave amplitudes in this analysis.  To estimate the
significance of each component, the fit is repeated with
the corresponding amplitude excluded and the
statistical significance is then determined by the changes of the
$-2\ln{\mathcal {L}}$ value with the number of degrees of freedom equal to twice
the number of extra parameters in the fit~\cite{PhysRevLett.115.072001}.
The statistical significances of all three components are found to be
larger than $24\sigma$. To check for an additional contribution, we
add an amplitude for the scalar meson $f_0(980)$, described by the
Flatt\'e function~\cite{Flatte976224} with the parameters fixed using
values from Ref.~\cite{Ablikim2005243}.  The corresponding statistical
significance is only $0.3\sigma$, and the contribution is therefore
neglected.

With the fitted values of $B_{0}^{P}$ = 2.685 $\pm$
0.006, $B_{1}^{P}$ = 1.740 $\pm$ 0.004, $B_{0}^{S}$ =
-39.09 $\pm$ 5.66, and $B_{1}^{S}$ = -39.18 $\pm$ 4.64, the
corresponding poles of $\rho$ and $\sigma$ are determined to be
$775.49 (\text{fixed}) - i(68.5\pm0.2)\;\text{MeV}$ and $(512\pm15)$ -
$i(188\pm12)\;\text{MeV}$, respectively, and are therefore in
reasonable agreement with the $\rho^\pm$ and $\sigma$ values from the
Particle Data Group (PDG)~\cite{PDG2014090001}.  The signal yields defined as the integrals
over the Dalitz plot of a single decay amplitude squared, the
detection efficiencies obtained from the MC sample weighted with each
amplitude and the branching fractions for each component are
summarized in Table~\ref{tab:EvtSwave}. In the
calculation, the number of $\jpsi$ is taken from
Refs.~\cite{Yanghx2014, Yanghxnew}, and the branching fraction for
$\jpsi\ra\gamma\etap$ and
$\pio\ra\gamma\gamma$ are taken from the
PDG~\cite{PDG2014090001}.

In order to compare with previous measurements which did not
consider the $P$ wave
contribution~\cite{Naik2009061801,Ablikim2012182001}, we also provide
the branching fraction of $\etap\ra\pip\pim\pio$ calculated with the
total number of observed signal events, which is presented in
Table~\ref{tab:EvtSwave}.

To check charge conjugation in the $P$-wave process, alternative fits
were performed with different magnitudes and phases for $\rhop$ and
$\rhom$. The result is consistent with charge symmetry:
$[\mathcal{B}(\etap\ra\rhop\pim)-\mathcal{B}(\etap\ra\rhom\pip)]/
[\mathcal{B}(\etap\ra\rhop\pim)+\mathcal{B}(\etap\ra\rhom\pip)]
=0.053\pm0.060(stat)\pm0.010(syst)$.

\begin{table}[!htbp]
\centering
 \caption{\label{tab:EvtSwave} Yields with statistical errors,
 detection efficiencies, and branching fractions for the studied $\etap$ decay modes,
 where the first errors are statistical, the second systematic, and the third model dependent.}
 \begin{tabular}{lcrc}\hline\hline
  	Decay mode        & 		Yield       & $\varepsilon$ (\%)&	 $\mathcal{B}$ ($10^{-4}$)	\\ \hline
    $\pip\pim\pio$    &  6067 $\pm$ 91  &	   25.3	          &  $35.91\pm0.54\pm 1.74$ \\
   $\pio\pio\pio$     &  2015 $\pm$ 47  &	   8.8	          &  $35.22\pm0.82\pm 2.54$ \\  \hline
$\rho^{\pm}\pi^{\mp}$ &  1231 $\pm$ 98  &    24.8           &  $7.44 \pm0.60\pm 1.26\pm 1.84$ \\
$(\pip\pim\pio)_S$    &  6580 $\pm$ 134 &	   26.2           &  $37.63\pm0.77\pm 2.22\pm 4.48$ \\ 
  \hline\hline
	\end{tabular}
\end{table}

As an alternative model, the Gounaris-Sakurai
parametrization~\cite{RhoPar} is used to describe the $\rho^\pm$
contribution with the mass and width fixed to the PDG
values~\cite{PDG2014090001}.  The $-\ln{\mathcal {L}}$ value is only
worse by 0.9.  In another check the $\pi$-$\pi$ $S$ wave for $\sigma$
is replaced with a relativistic Breit-Wigner function.  This fit also
provides a reasonable description of the data, and the $-\ln{\mathcal
  {L}}$ value only changes by 3.5.  The mass and width determined from
this fit are $(538\pm12)$ MeV/$c^2$ and $(363\pm20)$ MeV,
respectively, which are compatible with the pole position of the
$\pi$-$\pi$ elastic scattering amplitude.

Based on the symmetry imposed by Bose-Einstein statistics and
isospin~\cite{PhysRev133B1201,Bijnens2002qy}, the magnitude of the
nonresonant $S$ wave amplitude in $\etap\ra \pio\pio\pio$ is three
times that in $\etap\ra\pip\pim\pio$.  If this constraint is
introduced, the fitted yields are compatible with the unconstrained
result, while the change in $-\ln{\mathcal {L}}$ is 8.4, corresponding
to a statistical significance of $3.7\sigma$.

The differences of the branching fractions for the above tests
contribute to the systematic uncertainties, denoted as model and
constraint in Table~\ref{tab:syserr}, respectively.  In addition, the
following sources of the systematic uncertainty are considered:

The uncertainties in main drift chamber (MDC) tracking, photon selection and $\pio$
reconstruction efficiency (including photon detection efficiency) are
studied using a high purity control sample of $\jpsi\ra\rho\pi$.  The
differences between data and MC simulation are less than 1\% per
charged track, 1\% for the radiative photon and 2\% per $\pio$.

The uncertainties associated with kinematic fits are
studied using the control sample
$\jpsi\ra\gamma\eta\ra\gamma\pi\pi\pi$.  The preliminary selection
conditions for good charged tracks, good photons, and $\pio$
candidates are the same as those for
$\jpsi\ra\gamma\etap\ra\gamma\pi\pi\pi$.  The differences between data
and MC simulation for the requirements of
$\chi^{2}_{6C}(\gamma\pip\pim\pio)<25$ and
$\chi^{2}_{8C}(\gamma\pio\pio\pio)<70$ are determined as $1.7\%$
and $1.6\%$, respectively.

To investigate the uncertainties of the background determination,
alternative fits are performed on the background components one at a
time.  The peaking backgrounds $\etap\ra\gamma\rho$ and
$\etap\ra\pio\pio\eta$ are varied according to the errors of the
branching fraction for $\jpsi\ra\gamma\etap$ and the cascade decays in the 
PDG~\cite{PDG2014090001}.  The continuum background is varied
according to the uncertainties of the fits to the $\pi\pi\pi$ mass
spectra.  Different selection criteria for vetoing $\omega$ background
are also used.  The differences of the branching fractions with
respect to the default values are taken as the uncertainties
associated with backgrounds.

All the systematic uncertainties including the uncertainty from the
number of $J/\psi$ events and the branching fraction of $J/\psi
\ra\gamma\eta^\prime$ are summarized in Table~\ref{tab:syserr}, where
the total systematic uncertainty is given by the quadratic sum,
assuming all sources to be independent.

\begin{table}[!htbp]
 \caption{\label{tab:syserr} Summary of systematic uncertainties for the determination of
 			branching fractions for each component.}
 \begin{tabular}{lcccc}\hline\hline
    Source   &   $\rho^\pm\pi^\mp$  & $(\pip\pim\pio)_S$ & $\pip\pim\pio$ &  $\pio\pio\pio$ \\
     &   (\%) &   (\%) &   (\%) &   (\%) \\ \hline
		Constraint						        	& 15.9  	   & 3.3   & -       &  -     \\
		MDC tracking 	    				 			& 2     	   & 2	   & 2	     &  - 		 \\
		Radiative photon 	    			 		& 1     	   & 1	   & 1	     &  1 		 \\
		$\pio$ selection	    		      & 2          & 2	   & 2	     &  6 		 \\
		Kinematic fit					        	& 1.7   	   & 1.7   & 1.7     &  1.6   \\
		Background             	        & 3.0        & 1.4   & 1.2     &  1.3   \\
		Number of $\jpsi$  	            & 0.8        & 0.8	 & 0.8	   &  0.8   \\
		${\mathcal B}(\jpsi\ra\gamma\etap)$ & 3.1 	 & 3.1	 & 3.1	   &  3.1   \\ \hline
		Total 								         	& 16.9       & 5.9  & 4.9     &  7.2   \\ \hline\hline

		Model                				    & 24.7       & 11.9  & -       &  -     \\  
 \hline
 \end{tabular}
\end{table}

In summary, using a combined amplitude analysis of $\etap\ra\pip\pim\pio$ and $\etap\ra\pio\pio\pio$ decays, 
the $P$-wave contribution from $\rho^\pm$ is observed for the first time
with high statistical significance.
The pole position of $\rho^\pm$, $775.49 (\text{fixed})-i(68.5\pm0.2)\;\text{MeV}$, is consistent with
previous measurements, and the branching fraction ${\mathcal B}(\etap\ra\rho^{\pm}\pi^{\mp})$
is determined to be $(7.44\pm0.60\pm1.26\pm1.84)\times 10^{-4}$.

In addition to the nonresonant $S$ wave, the resonant $\pi$-$\pi$
$S$ wave with a pole at $(512\pm15)-i(188\pm12)$ MeV, 
interpreted as the broad $\sigma$ meson, plays an essential role
in the $\etap\ra\pi\pi\pi$ decays.
Because of the large interference between
nonresonant and resonant $S$ waves, only the sum is used to describe the $S$-wave
contribution, and the branching fractions are determined to be
${\mathcal B}(\etap\ra\pip\pim\pio)_S=(37.63\pm 0.77\pm2.22\pm4.48)\times10^{-4}$ and
${\mathcal B}(\etap\ra\pio\pio\pio)=(35.22\pm0.82\pm2.54)\times10^{-4}$, respectively.
The branching fractions of $\etap\ra\pip\pim\pio$ and
$\etap\ra\pio\pio\pio$ are in good agreement with and supersede the
previous BESIII measurements~\cite{Ablikim2012182001}.  The value for
${\mathcal B}(\etap\ra\pio\pio\pio)$ is two times larger than that
from GAMS
[$(16.0\pm3.2)\times10^{-4}$]~\cite{Alde1987603}.
The significant resonant $S$-wave contribution also provides a
reasonable explanation for the negative slope parameter of the
$\etap\ra\pio\pio\pio$ Dalitz plot~\cite{PhysRevD92012014}.  The ratio
of the branching fractions between the $S$-wave components
$\mathcal{B}(\etap\ra\pio\pio\pio)/\mathcal{B}(\etap\ra\pip\pim\pio)_S$
is determined to be $0.94\pm0.03\pm0.13$, where the common systematic
uncertainties cancel out.  With the branching
fractions of $\etap\ra\pi\pi\eta$ taken from the PDG~\cite{PDG2014090001},
$r_{\pm}$ and $r_{0}$ are now calculated to be
$(8.77\pm1.19)\times10^{-3}$ and $(15.86\pm1.33)\times10^{-3}$,
respectively. While the previous values based on the PDG~\cite{PDG2014090001} are
$(8.86\pm0.94)\times10^{-3}$ and $(9.64\pm0.97)\times10^{-3}$,
respectively.

The observed substantial $P$- and $S$-wave resonant contributions have
to be properly considered by theory before attempting to determine
light quark masses from $r_{\pm}$ and $r_0$. In
particular, one of the previously most comprehensive analyses of hadronic
decays of $\eta$ and $\etap$ mesons relied on $r_0$,
which is now two times larger, and $r_\pm$ was not
known~\cite{Borasoy200641}. 
Further progress will depend on the development of dispersive approaches
such as Refs.~\cite{Colangelo:2009db,Schneider:2010hs,Kampf:2011wr,Guo:2015zqa} for $\etap$ hadronic decays.

\begin{acknowledgments}
The BESIII Collaboration thanks the staff of BEPCII and the IHEP computing center for their strong support.
This work is supported in part by
National Key Basic Research Program of China under Contract No. 2015CB856700,
National Natural Science Foundation of China (NSFC) under Contracts No. 11675184, No. 11125525,
No. 11235011, No. 11322544, No. 11335008, and No. 11425524,
the Chinese Academy of Sciences (CAS) Large-Scale Scientific Facility Program,
Joint Large-Scale Scientific Facility Funds of the NSFC and CAS under Contracts
No. 11179007, No. U1232201, and No. U1332201,
Youth Science Foundation of China under Contract No. Y5118T005C,
CAS under Contracts No. KJCX2-YW-N29 and No. KJCX2-YW-N45,
100 Talents Program of CAS, INPAC and Shanghai Key Laboratory for Particle Physics and Cosmology,
German Research Foundation DFG under Contract No. Collaborative Research Center CRC-1044,
Instituto Nazionale di Fisica Nucleare, Italy,
Ministry of Development of Turkey under Contract No. DPT2006K-120470,
Russian Foundation for Basic Research under Contract No. 14-07-91152,
U. S. Department of Energy under Contracts No. DE-FG02-04ER41291, No. DE-FG02-05ER41374,
No. DE-FG02-94ER40823, and No. DESC0010118,
U. S. National Science Foundation,
University of Groningen (RuG) and the Helmholtzzentrum fuer Schwerionenforschung GmbH (GSI), Darmstadt,
and WCU Program of National Research Foundation of Korea under Contract No. R32-2008-000-10155-0.
\end{acknowledgments}

\bibliographystyle{apsrev}
\bibliography{etap23pi}

\end{document}